\documentclass[aps,showpacs,letterpaper,twocolumn,12point]{revtex4-1}

\usepackage{graphicx} 
\usepackage{amsmath}
\usepackage{color}      

\usepackage{amstext}
\usepackage{graphics}
\usepackage{amssymb}

\newcommand \be{\begin{equation}}
\newcommand \ee{\end{equation}}
\newcommand \bea{\begin{eqnarray}}
\newcommand \eea{\end{eqnarray}}
\newcommand \bse{\begin{subequations}}
\newcommand \ese{\end{subequations}}
\newcommand \nn{\nonumber}

\begin{document}

\title{A two-dimensional lattice of blue detuned atom traps using a projected Gaussian beam array}

\author{M. J. Piotrowicz, M. Lichtman, K. Maller}

\author{G. Li}
\altaffiliation{Present address: Institute of Opto-Electronics, Shanxi University
92 Wucheng Road, Taiyuan 030006, Shanxi Province, China}
\author{S. Zhang}
\altaffiliation{Present address: KLA Tencor, 
One Technology Drive
Milpitas, Ca. 95035}
\author{L. Isenhower,  and M. Saffman}

\affiliation{
Department of Physics, University of Wisconsin, 1150 University Avenue, Madison, Wisconsin 53706
}

\begin{abstract}We describe a new type of blue detuned optical lattice for atom trapping which is intrinsically two dimensional, while providing 
three-dimensional atom localization. The lattice is insensitive to optical phase fluctuations since it does not depend on  field interference between  distinct optical beams.    The array is created using a novel arrangement of weakly overlapping Gaussian beams that creates a  two-dimensional array of dark traps which are suitable for magic trapping of ground and Rydberg states. We analyze the spatial localization that can be achieved and demonstrate trapping and detection of single Cs atoms in 6 and  49 site two-dimensional arrays.  \end{abstract}

\pacs{37.10.Gh,37.10.Jk,03.67.-a}

\date{\today}

\maketitle

\section{Introduction}

Arrays of neutral atom qubits  in optical traps
are being actively developed for implementing multi-qubit quantum
information processing (QIP) devices\cite{Meschede2006,Saffman2010,Negretti2011,Schlosser2011}.
Far detuned optical traps provide strong confinement with low photon scattering rates and low decoherence\cite{Cline1994}. Red detuned traps
localize atoms at a maximum of the optical intensity which leads to higher photon scattering rates and light shifts on atomic levels which are used for qubit encoding and control. Blue detuned traps confine atoms at a local minimum of the optical intensity. This reduces photon scattering and light shifts, and is of particular interest for experiments using Rydberg atom excitation since blue detuned configurations allow for simultaneous trapping of both ground and Rydberg excited states\cite{SZhang2011}. This capability will be important for future scalable QIP devices based on Rydberg state mediated quantum gates\cite{Wilk2010,Isenhower2010,Zhang2010} as well as adiabatic approaches based on Rydberg dressing\cite{Keating2013} or dissipative interactions\cite{Carr2013} which require long term occupancy of Rydberg states.

Projected arrays of dipole traps have been demonstrated using either
microlenses\cite{Dumke2002}, holographic methods\cite{Bergamini2004}, or diffractive optics\cite{Knoernschild2010,Gillen-Christandl2010}.
 Several experiments in recent years have demonstrated loading of single atoms into small arrays of optical traps\cite{Bergamini2004,Knoernschild2010} and into larger optical lattices using either stochastic loading\cite{Nelson2007} or Bose-Einstein condensate to Mott insulator
techniques\cite{Bakr2009,Sherson2010,Weitenberg2011}. For QIP applications we would like the trap array to have the following characteristics.
It should be scalable to a large number of trapping sites, two-dimensional to minimize crosstalk from neighboring planes of trapped atoms, stable against trap position drifts due to optical phase fluctuations, and, particularly for
 experiments with Rydberg atoms\cite{Saffman2010}, we wish to use blue detuned traps.

Most optical lattices use interference of beams that are counter-propagating, or co-propagating at a small angle, to create the trap array\cite{Petsas1994}. With this approach the positions of the trap sites are directly sensitive to optical path length drifts in the apparatus causing differential phase shifts between beams. Although active stabilization is possible\cite{Weidemuller1998} this has not been demonstrated in single atom experiments. Alternatively one of the sites can be used to monitor lattice drifts\cite{Weitenberg2011}.
A diffractive method was demonstrated in \cite{Bakr2009} which suppresses phase sensitivity of the lattice sites. That method provides a two-dimensional trap array with localization out of the plane being provided by a separate orthogonally propagating beam.

\begin{figure}[!t]
 \centering
  \begin{minipage}[c]{8.5cm}
    \centering
 \includegraphics[width=8.5cm]{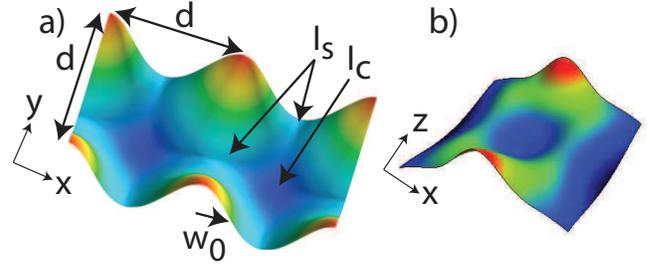}
\end{minipage}\caption{(color online)  Intensity distribution of Gaussian beam array in the $x-y$ (a) and $x-z$ (b) planes. The beams propagate along the $z$ axis giving an array of traps lying in the $x-y$ plane. The trap array has periodicity $d$ and forms atom traps at the center of each four beam plaquette where the intensity is $I_c$. Lateral confinement in the $x-y$ plane is provided by the saddles with intensity $I_s$. Localization normal to the $x-y$  plane along $z$ is provided by diffractive spreading of the beams. 
 }
\label{fig.GBAunitcell}
\end{figure}

Here we propose and demonstrate a new approach that is scalable to many sites, creates an intrinsically two-dimensional array of traps which localize the atoms in all three dimensions, and is blue detuned for use with Rydberg atoms. This work builds on recent  experience with atom trapping in  blue detuned bottle beam traps 
(BBTs)\cite{Isenhower2009,Xu2010,Li2012,Ivanov2013}. A possible approach would be to create multiple copies of a BBT using a diffractive beam splitter. Such an array confines atoms with two trap walls between each trapped atom. This leads to a lower density of sites than we would like. Instead we use a diffractive beam splitter to create a weakly overlapping Gaussian beam array (GBA). The atoms are localized in the intensity minima between beams while the overlap regions create saddle potentials which laterally trap the atoms, as shown in Fig. \ref{fig.GBAunitcell}. Localization out of the plane is provided by diffractive spreading of the beams.

The rest of this paper is organized as follows. In Sec. \ref{sec.design} we describe the design of the GBA and present two different versions which we refer to as half- and full-incoherent. We also compare the performance in terms of trap depth and localization with a conventional optical lattice.  In Sec. \ref{sec.optics} we describe the optical system used to create the array and in Sec. \ref{sec.atoms} we show that single atoms can be effectively trapped. We conclude with an outlook in Sec. \ref{sec.conclusion}.

\section{Gaussian beam array design}
\label{sec.design}

In this section we present the design and analysis of an array of blue detuned traps based on a weakly overlapping Gaussian beam array (GBA).
The geometry is shown in Fig. \ref{fig.GBAunitcell}. Each beam has a waist parameter $w_0$ (radius where the intensity is  $1/e^2$ of the maximum) and the array  periodicity is $d$.  We will frequently use the ratio $s=d/w_0$ to characterize the array. We use the term weakly overlapping to describe the situation where $d>w_0$ and the ratio is $s\sim 2$. For this value of $s$ the overlap between neighboring beams is significant, and indeed defines the trap sites, while the overlap between beams separated by larger distances is negligible. 

In order to suppress coherent interference between  neighboring beams we analyze two types of array. In the first, which we call half-incoherent, neighboring beams have orthogonal polarizations so we can add their intensities when calculating the trap depth.  This statement is based on the assumption that
the beams are far detuned from the nearest atomic resonance so that vector and tensor contributions to the ground state polarizability are negligible.
The remaining field interference terms are between beams on the diagonal of a unit cell. Their separation is $\sqrt2 d$, whereas neighboring beams are separated by only $d$ so the interference term along the diagonal is strongly suppressed compared to neighboring beams separated by $d$. Nevertheless the sensitivity to phase variations between beams can lead to 
variations in the trap intensity at the center of each unit cell.

The second type of array which we call full-incoherent effectively removes the residual phase sensitivity of the half-incoherent array.  In this case we use a combination of orthogonal polarizations and different laser frequencies so that there is no field interference between neighbors, or between diagonal neighbors. As long as the frequency difference is large compared to the trap vibrational frequencies, we can treat the array potential as being due to the incoherent sum of the beams.  
The remaining field interference effects are due to beams separated by $2d$ or more, and for our parameters these terms are negligible.

For both the half-incoherent and full-incoherent arrays the use of different polarizations and frequencies also serves an additional purpose. If the entire array was due to a uniformly polarized coherent field the Talbot effect would lead to multiple copies of the array along the normal $z$ axis. This would allow for trapping in multiple planes and we would not have a two-dimensional array of traps. The half- and full-incoherent designs effectively suppress the Talbot effect and we get a single plane of traps.

\subsection{Half-incoherent array}
\label{sec.halfIncoherent}

\begin{figure}[!t]
 \centering
  \begin{minipage}[c]{8.5cm}
    \centering
 \includegraphics[width=8.5cm]{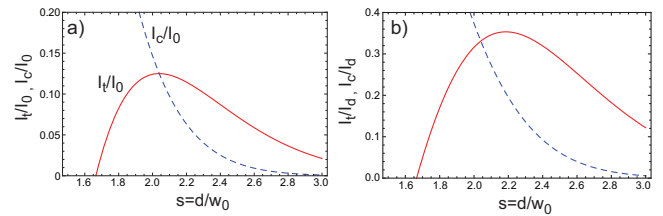}
\end{minipage}\caption{(color online)  Normalized trapping depth (solid red curve) and intensity at trap center (dashed blue curve) versus normalized array period for half-incoherent array. Variation at constant peak intensity $I_0$ in a), and variation at constant average intensity $I_d$ in b). }
\label{fig.gbarraydepthhi}
\end{figure}

Consider a unit cell with neighboring beams having orthogonal polarization states as shown in Fig. \ref{fig.GBAunitcell}.
 At the saddle point along each side of the unit cell the intensity is  approximately
\be
I_{\rm s} = 2 I_0 e^{-2 (d/2)^2/w_0^2}= 2 I_0 e^{- d^2/2w_0^2},
\label{eq.GBAIs}
\ee
where $I_0$ is the peak intensity of one beam.
If diagonally opposite beams are in phase (this is the worst case giving the smallest trapping potential)
 the intensity  at the center of the unit cell where an atom is trapped is approximately
\be
I_{\rm c} = 2 I_0 \left(2e^{-(d/\sqrt2)^2/w_0^2}\right)^2
= 8I_0 e^{-d^2/w_0^2}.
\label{eq.GBAIc}
\ee
 In Eqs.  (\ref{eq.GBAIs},\ref{eq.GBAIc}) we neglect contributions from further away beams in neighboring unit cells; for the saddle intensity we only account for the two nearest beams and for the center intensity we account for all 4 beams in one unit cell.

The trap depth is proportional to the difference of these two intensities which is
\begin{eqnarray}
I_{\rm t} &=& I_{\rm s}-I_{\rm c}\nonumber\\
&=&I_0 \times 2 e^{- s^2/2}\left( 1 - 4 e^{- s^2/2}\right).
\label{eq.gbarraythi}
\end{eqnarray}
Figure \ref{fig.gbarraydepthhi}a) shows the trap depth as a function of $s$. The trap depth has a maximum at $s_0=(2\ln8)^{1/2}\simeq2.04
$. Using $s=s_0$ we find $I_s=I_0/4$, $I_c=I_0/8,$ and $I_t = I_0/8.$

We will be most interested in the trap depth as a function of $w_0$ for fixed optical power and fixed lattice period $d$. Defining the average intensity in a unit cell of area $d^2$ as $I_d=\frac{P}{d^2}=\frac{\pi w_0^2 I_0}{2d^2}=\frac{\pi I_0}{2 s^2}$ the trap  intensity is
\be
I_{\rm t}=  I_d\times  \frac{4s^2 e^{- s^2/2}}{\pi}\left( 1 - 4 e^{- s^2/2}\right).
\ee
Figure \ref{fig.gbarraydepthhi}b) shows the trap depth
which reaches a maximum of $I_t/I_d=0.35$ at  $s=2.19$. Note that the calculated trap depth in Fig. \ref{fig.gbarraydepthhi} is based on a worst case assumption. If the diagonal beams were out of phase we would get $I_c=0$ and the peak trap depth would be about twice bigger. It can also be verified that contributions from further away beams have a negligible impact on the plots in Fig. \ref{fig.gbarraydepthhi}.

In addition to the trap depth it is important to know the spatial localization and oscillation frequencies. Using the approximation of the potentials at the trap center given in \cite{SZhang2011} we get the effective spring constants
\bse\bea
\kappa_x &=&\frac{32 |U_d|}{\pi d^2}s^4 (s^2-2)e^{-s^2} \\
\kappa_y &=&\kappa_x \\
\kappa_z &=&\frac{32 \lambda^2|U_d|}{\pi^3  d^4}s^6(s^2-1)e^{-s^2}
\eea
\label{eq.GBAkappahi}
\ese
with $U_d=\frac{\alpha}{2\epsilon_0 c} I_d$, $\alpha$ is the atomic polarizability in SI units, and $\lambda$ is the wavelength of the trapping light. The $x$ axis is directed from the trap center
towards a neighboring side. The corresponding oscillation frequencies are $\omega=\sqrt{\kappa/m_a}$
with $m_a$ the atomic mass.

\begin{figure}[!t]
 \centering
  \begin{minipage}[c]{8.5cm}
    \centering
 \includegraphics[width=8.5cm]{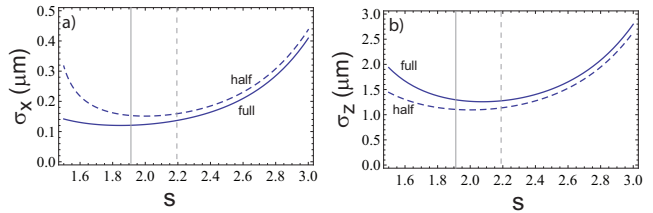}
\end{minipage}\caption{(color online)  Standard deviation of $x$ position (a) and $z$ position (b) for a half-incoherent array (dashed curves) and full-incoherent array (solid curves). The vertical lines are at the $s$ values for which the trap depths are maximized.  Parameters are $\lambda=0.78~\mu\rm m$, $d=3.6~\mu\rm m$, $T=10~\mu\rm K$, and $U_t=k_B\times 300 ~\mu\rm K$.}
\label{fig.GBAlocalize}
\end{figure}

The time averaged position variances are found from $\frac{1}{2}\kappa_j \sigma_j^2=\frac{1}{2}\kappa_j \langle r_j^2\rangle=\frac{1}{2} k_B T$ with $T$ the atomic temperature. They are
\bse\bea
\sigma_x^2 &=&\frac{\pi k_B T}{ 32|U_d|}\frac{e^{s^2}}{s^2(s^2-2) }w_0^2=\sigma_{x0}^2\frac{e^{s^2}}{s^4(s^2-2) },\\
\sigma_y^2 &=&\sigma_x^2,\\
\sigma_z^2&=& \frac{\pi k_B T}{ 32|U_d|}\frac{e^{s^2}}{s^2(s^2- 1)}L_R^2=\sigma_{z0}^2\frac{e^{s^2}}{s^6(s^2- 1)},
\eea
\label{eq.sigmaxzhi}
\ese
with $\sigma_{x0}=\left(\frac{\pi d^2 k_B T }{ 32|U_d|}\right)^{1/2}$, $\sigma_{z0}=\left(\frac{\pi^3 d^4k_B T}{ 32\lambda^2|U_d|}\right)^{1/2}$.
The optimal confinement values are $\sigma_x=1.31\, \sigma_{x0}$  at
$s=2$ and  $\sigma_z=0.53\, \sigma_{z0}$  at
$s=\left( \frac{5+\sqrt{13}}{2}\right)^{1/2}\simeq 2.07$. Fig. \ref{fig.GBAlocalize} shows the dependence of confinement on the parameter $s$.

\subsection{Full-incoherent array}

If we use two laser frequencies we can arrange for all neighboring beams, both along an edge, and across a diagonal of a unit cell to combine incoherently. An optical layout which implements this will be shown in Sec. \ref{sec.optics}.
For this arrangement the saddle and center intensities are
\be
I_{\rm s} = 2 I_0 e^{- s^2/2},
\label{eq.GBAIs2}
\ee
and
\be
I_{\rm c} = 4 I_0 e^{- 2(d/\sqrt2)^2/w_0^2}
= 4I_0 e^{-s^2}.
\label{eq.GBAIc2}
\ee
Comparing with (\ref{eq.GBAIs}, \ref{eq.GBAIc}) we see that the saddle intensity is unchanged, but the intensity at the center is reduced by a factor of two. Reduction of the center intensity increases the trap depth, and also eliminates the phase dependence  which would lead to unwanted structure near the center of the trap.

\begin{figure}[!t]
 \centering
  \begin{minipage}[c]{8.5cm}
    \centering
 \includegraphics[width=5.5cm]{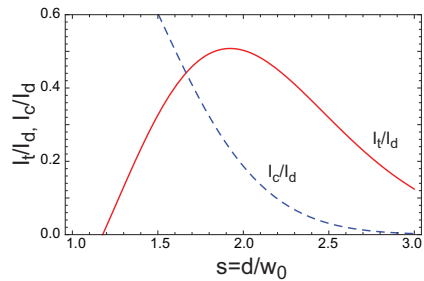}
\end{minipage}\caption{(color online)  Normalized trapping depth (solid curve) and intensity at trap center (dashed curve) versus normalized array period for full-incoherent array. The plot assumes constant average intensity $I_d$.  }
\label{fig.gbarraydepth}
\end{figure}

The trap depth is
\bea
I_{\rm t} &=&I_0\times 2 e^{- s^2/2}\left( 1 - 2 e^{- s^2/2}\right)\nn\\
&=&I_d\times \frac{4 s^2 e^{- s^2/2}}{\pi} \left( 1 - 2 e^{- s^2/2}\right)
\label{eq.gbarrayt}
\eea
Figure \ref{fig.gbarraydepth} shows the trap depth as a function of $s$. The maximum of $I_t/I_d=0.51$ occurs at $s=1.92.$
For the same average intensity the full-incoherent array has about 30\% larger trap depth than the half-incoherent array.

Following the same steps as for the half-incoherent array we find for the
spring constants and spatial localization
\bse\bea
\kappa_x&=& \frac{32 |U_d| }{\pi d^2}s^4(s^2-1)e^{-s^2},\\
\kappa_y&=&\kappa_x\\
\kappa_z &=&\frac{16 \lambda^2|U_d|}{\pi^3  d^4}s^8e^{-s^2}
\eea\ese
and
\bse\bea
\sigma_x^2 &=&\sigma_{x0}^2\frac{e^{s^2}}{s^4(s^2-1) },\\
\sigma_y^2 &=&\sigma_x^2 \\
\sigma_z^2 &=& \sigma_{z0}^2\frac{2e^{s^2}}{s^8}.
\eea
\label{eq.sigmaxzfi}
\ese
The optimal confinement values are $\sigma_x=1.04\, \sigma_{x0}$  at
$s=\sqrt{2+\sqrt2}\simeq1.85$ and  $\sigma_z=0.65\, \sigma_{z0}$  at
$s=2.$

Comparing the half- and full-incoherent arrays we see that the full-incoherent case has better trap depth and transverse localization for the same optical power.
The axial localization is worse for the full-incoherent array, but this is a less important figure of merit than the transverse localization when addressing beams propagate along $z$. We conclude that the full-incoherent array should give somewhat better performance for qubit control, and also is less sensitive to parasitic light scattering and optical imperfections.

\subsection{Comparison with  optical lattice}

The spatial average of the intensity in a  unit cell $I_d$ is  related to the effective trapping intensity $I_t$ by
\bea
I_t^{\rm hi} &=&I_d\times  \frac{4s^2 e^{-s^2/2}}{\pi}\left(1-4 e^{-s^2/2} \right)\\
I_t^{\rm fi} &=&I_d\times  \frac{4s^2 e^{-s^2/2}}{\pi}\left(1-2 e^{-s^2/2} \right),
\eea
where hi, fi stand for half- and full-incoherent.
The maximum optical efficiency is $(I_t/I_d)_{\rm hi}=0.35$ and
$(I_t/I_d)_{\rm fi}=0.51.$
These ratios can be compared with a traditional optical lattice formed by interfering plane waves. The comparison depends on the dimensionality and type of optical lattice. For the simplest case of a one dimensional lattice  the effective trapping
intensity is the difference of the maximum and minimum intensities which is twice the average intensity for a sinusoidal lattice. Thus the full-incoherent Gaussian beam array implementation has a relative efficiency of at best $0.51/2 = 25.5\%$.

The GBA appears more favorable when we consider the performance in three dimensions.
 We will consider two types of coherent lattice. A standard counterpropagating geometry uses three beams, each retroreflected to create a $d=\lambda/2$ lattice period in all three dimensions. The power required is three times $P_1$, the power of each beam, so the intensity averaged over a unit cell is $I_d= 3 P_1/(\lambda/2)^2=12 P_1/\lambda^2.$ The trap depth is $I_t=4 I_1$ so
$I_t/I_d= \lambda^2 I_1/3 P_1$, and putting $I_1=P_1/(\lambda/2)^2$ we get
$I_t/I_d=4/3$. The GBA has a relative efficiency of $.51/(4/3)=39\%$.
It is also possible to create longer period lattices by interfering pairs of beams at an acute angle $\theta<\pi$ as in \cite{Nelson2007}. In this case $I_t=4 I_1=4 P_1/d^2$, $I_d=6P_1/d^2$ and $I_t/I_d=2/3.$  The GBA has a relative efficiency of $.51/(2/3)=78\%$.

We see that the GBA has a lower efficiency than a standard optical lattice.
The benefit is the absence of phase sensitivity as regards the position of the trap sites as well as the ability to project the lattice onto the atomic sample using optical access from a single side.

\section{Optical implementation}
\label{sec.optics}

A straightforward approach to creating an array of Gaussian beams is to start with a single laser beam and use a Dammann grating \cite{Dammann1971} to replicate the beam. Gratings that can create several hundred equal intensity beams are readily available commercially. Unfortunately a single Dammann grating does not work well for creating an array of overlapping beams. The diffractive spreading angle of a Gaussian beam of waist $w_0$ is
$\theta_d=\lambda/\pi w_0$. The angular separation of the beams from a grating of period $\Lambda$ is $\theta_g=\lambda/\Lambda$.  In order to achieve $s=d/w_0\simeq 2,$ which we identified in the previous section as the optimal spacing, we need $\theta_g/\theta_d\simeq 2$. However, this implies that
$\theta_g/\theta_d=\pi w_0/\Lambda\simeq 2$ or $\Lambda\simeq \frac{\pi}{2}w_0$. In other words the grating period is comparable to the waist of the Gaussian beam illuminating the grating. In this regime standard Dammann gratings do not perform well and lead to large distortions of the
diffracted spot array. Our tests show that the beam quality after the grating is good when $s\gtrsim 4$.

\begin{figure}[!t]
 \centering
  \begin{minipage}[c]{8.5cm}
    \centering
 \includegraphics[width=8.5cm]{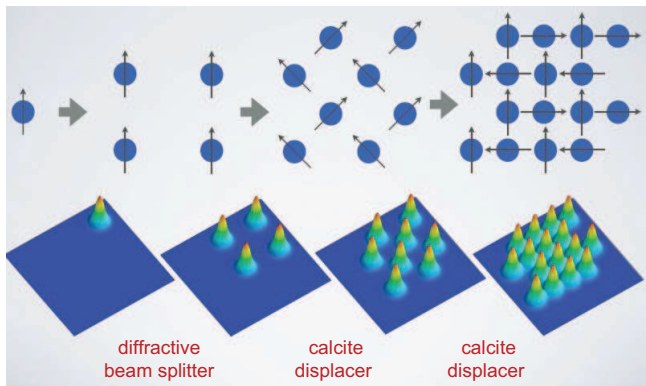}
\end{minipage}\caption{(color online)  Making a half-incoherent array with a Dammann grating and calcite displacers. The illustration shows creation of a 16 beam array with 6 trapping sites. The arrows indicate the direction of linear polarization of each beam. }
\label{fig.gbarray}
\end{figure}

While it should be possible to design a custom grating that works well for this application we have instead used a Dammann grating with $\Lambda\simeq \pi w_0$ giving $s=4$ followed by beam displacement optics as shown in Fig. \ref{fig.gbarray} to reduce the beam spacing to $s=2$. The calcite beam displacement elements serve the additional function of giving neighboring beams orthogonal polarizations for the  array designs described above.

\begin{figure}[!t]
 \centering
  \begin{minipage}[c]{8.5cm}
    \centering
 \includegraphics[width=8.5cm]{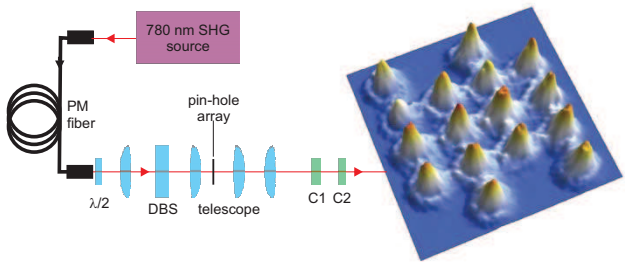}
\end{minipage}\caption{(color online) Optical system for creating a half-incoherent GBA with 6 trapping sites. The intensity image shows the array before focusing onto the atoms with parameters   $\lambda=0.78~\mu\rm m$, $d=250~\mu\rm m$, $w_0=120~\mu\rm m$, and $s=2.1$. The 780 nm source is based on a frequency doubled single frequency 1560 nm laser. }
\label{fig.gbarrayExperiment}
\end{figure}

We have implemented a half-incoherent GBA as  shown in Fig. \ref{fig.gbarrayExperiment}. A single TEM$_{00}$ Gaussian beam is divided on a Dammann grating which we refer to as a diffractive beam splitter (DBS) (Holo/Or MS-248-X-Y-A). The grating is placed in the front focal plane of a lens to create an array of four parallel beams with separation $d=500~\mu\rm m$ and
$s\simeq 4.2$. The pin-hole array suppresses all non-first order beams from the DBS.
 This is then followed by
two pieces of calcite. The first, thicker piece C1 is cut to give a lateral displacement of $500/\sqrt2=354~\mu\rm m$. This is then followed with a second, thinner  calcite C2 rotated by 45$^\circ$ relative to C1, which gives a displacement $500/2=250~\mu\rm m$. The net result is an array with $d=250~\mu\rm m$ and
$s\simeq 2.1$ with neighboring beams having orthogonal linear polarizations. An implementation of this design to create 16 beams and 6 trap sites is shown in Fig. \ref{fig.gbarrayExperiment}. The 16 beam array is then imaged onto a cloud of cold Cs atoms in a pyrex cell with a  multielement NA=0.4 lens that is designed to compensate for cell wall aberrations to give diffraction limited focusing. The lattice spacing at the atoms is $d=3.8~\mu\rm m$.

\begin{figure}[!t]
 \centering
  \begin{minipage}[c]{8.5cm}
    \centering
 \includegraphics[width=8.5cm]{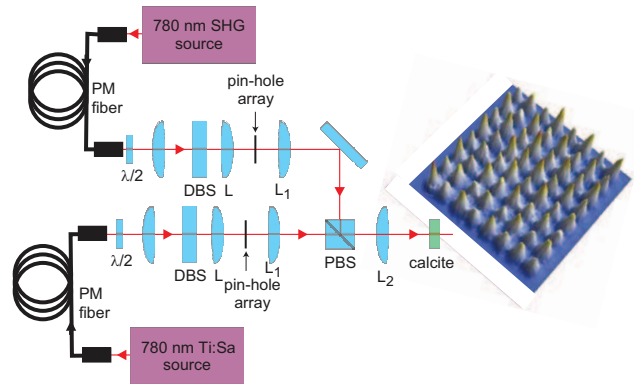}
\end{minipage}\caption{(color online)  Experimental setup for creating a full-incoherent GBA with 49 trapping sites. Two $4\times 4$ arrays created by diffractive elements (DBS) are combined by polarizing beam-splitters (PBS). Lenses $L_1$ and $L_2$ create a 1:1 telescope to image the arrays onto the calcite that combined them into a 64 beam array. Pin-hole arrays are placed in the foci of lenses $L$ to block unwanted zeroth and higher orders of diffraction from the DBS. The spatial parameters of the resulting array are the same as in Fig. \ref{fig.gbarrayExperiment}. }
\label{fig.GBAsetupFI}
\end{figure}

We have also implemented the full-incoherent array design, where the nearest neighbor beams  come from two different laser sources (or are shifted in frequency by acousto-optic modulators) and the next nearest neighbors have different polarizations. The scheme of the optical realization is presented in Fig. \ref{fig.GBAsetupFI}. Two separated arms  create $4\times 4$ Gaussian arrays that are combined using a polarizing beam splitter (PBS). Finally, this array is shifted by a single calcite to create a 64 beam array with 49 trapping sites. The other parameters for the trap array are similar to the half-incoherent case described above.

\section{Atom trapping demonstration}
\label{sec.atoms}

We have demonstrated that both the half- and full-incoherent arrays described above are suitable for trapping of cold Cs atoms. We use a double magneto-optical trap (MOT) apparatus with a 2D MOT feeding a  3D MOT in a differentially pumped pyrex vacuum cell. We 
 load the GBA from the 3D MOT which is based on  a standard 6 beam configuration.  The MOT is cooled to $10-20~\mu\rm K$  with  10 ms of polarization gradient cooling  (PGC) giving number densities $\sim 2\times 10^9$~cm$^{-1}$.  The 780~nm trapping light is switched on at the beginning of the PGC phase when the atomic density is highest and at the end of the loading phase all MOT and PGC light is  switched off to allow the atoms that are not trapped to fall away. The MOT and GBA loading takes about 0.7 s. Trapped atoms are detected by turning on the PGC light and imaging the scattered fluorescence onto an
electron multiplying charge coupled device (EMCCD) camera. The 852 nm fluorescence is collected through the same lens used to project the GBA onto the atoms and separated with a dichroic filter from the 780 nm trapping light. Typical detection parameters are detuning of $-39~{\rm MHz} =- 7.5~\rm \gamma$ from the $6s_{1/2},f=4\leftrightarrow 6p_{3/2}, f=5$ transition
 with $I\simeq2.7 I_{\rm sat}$, $I_{\rm sat}$ is the saturation intensity, and scattered light is collected for 50~ms. EMCCD exposure times as short as 5 ms are sufficient to resolve single atom signals. 

\subsection{Half-incoherent array}

We  first describe results using the 6 site half-incoherent array.  
The trap depth is calculated from  Eq. (\ref{eq.gbarraythi}). For Cs atoms at a trap light wavelength of 780 nm the scalar polarizability is $\alpha_0^{\rm cgs} = -240\times 10^{-24}~\rm cm^3$. We transmit 3 W of 780 nm light through a single mode polarization maintaining fiber. The optical efficiency from the fiber end to the atoms including the array generation and subsequent relay and focusing optics is about 50\%. With a power of 1.5 W at the atoms divided into 16 beams we achieve 
trap depths of  $\sim 830~\mu$K in an array with $d=3.8~\mu\rm m$ and $s=2.1$. The array period is larger than is common in many optical lattice experiments and is chosen to be compatible with magic trapping of high $n$ Rydberg states\cite{SZhang2011}.
The close to $4~\mu\rm m$ trap spacing also facilitates addressing of single sites using laser beams focused to few micron waists\cite{Knoernschild2010}. 

\begin{figure}[!t]
 \centering
  \begin{minipage}[c]{8.5cm}
    \centering
 \includegraphics[width=8.5cm]{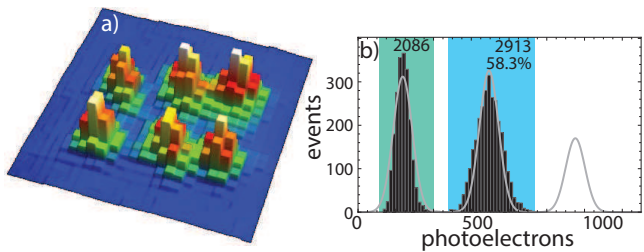}
\end{minipage}\caption{(color online)  Fluorescence image a) and atom number histogram b) in the 6 site half-incoherent array.
The image is an average of 105 out of 5000 events where all 6 sites loaded a single atom. Each pixel is $0.63\times 0.63 ~\mu\rm m^2$. The atom number histogram has 5000 events taken from one of the six sites showing clear separation of the 0 atom background from the 1 atom peak. The solid line is a Poissonian model  fitted to the 1 atom peak.  }
\label{fig.histogram}
\end{figure}

Figure  \ref{fig.histogram} shows fluorescence images of the trapped atoms and a histogram of single atom events.  In the histogram the single atom signal is clearly seen with the loading rate varying from 50\% to 60\% between sites.
The average loading rate is $52.5\%$ which is sub-Poissonian, and also slightly above that expected from light assisted two-body loss, or collisional blockade\cite{Schlosser2002}.  We do not observe any events with two or more atoms in the trap, although we cannot exclude the possibility that we load two atoms, but lose them rapidly during the fluorescence imaging exposure. 

\begin{figure}[!t]
 \centering
  \begin{minipage}[c]{.4\textwidth}
    \centering
 \includegraphics[width=\textwidth]{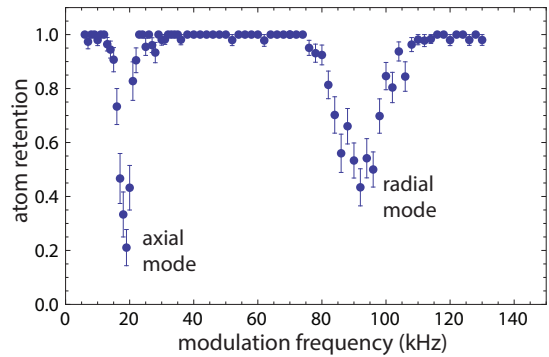}
\end{minipage}\caption{(color online)  Parametric heating measurement of trap frequencies.}
\label{fig.trapfreq}
\end{figure}

To compare the trap depth with  theoretical calculations we measured the trap oscillation frequencies. Equations (\ref{eq.GBAkappahi}) give the frequencies as $\omega_x=2\pi\times 39$~kHz and $\omega_z=2\pi\times 6.4$~kHz.
 To measure the trap oscillations the intensity of the trapping light was modulated at frequency $f$. When the frequency matches $2 \omega_0 / 2\pi$ the atoms will be heated and leave the trap. After  confirming the initial presence of an atom in the trap site the modulation was applied for 100~ms for $f<35 ~\rm kHz$ and for 5 ms for $f>35~\rm kHz$. 
Then we took a second  image to measure the retention of the atom in the trap. For each modulation frequency the measurement was repeated 200 times. The observed spectrum for one of the traps is presented in Fig. \ref{fig.trapfreq}. 

Two resonances can be easily identified: one around 
18~kHz corresponding to $\omega_z/2\pi$=9~kHz (axial frequency) and a broader resonance centered at 90~kHz ($\omega_x/2\pi$=45~kHz). The radial frequency agrees to about 10\% with that calculated based on our known trap geometry and optical power. 
The axial frequency is about 50\% higher than expected. We have observed that individual focused beams in the array do not have an ideal Gaussian profile and diverge faster than for a Gaussian with the same waist parameter, i.e. the beam quality 
factor is $M^2>1$. This fast divergence has a  minimal effect on the radial trap frequencies but will increase the axial frequencies which is consistent with our observations.

For  future experiments with qubits we wish the  traps to be well aligned on a regular grid. Fits to 
 the image in Fig. \ref{fig.histogram} reveal an average spacing of $d=3.90~\mu\rm m$ close to the expected $3.8~\mu\rm m$. 
We find a maximum deviation of the atomic centroids from a regular grid of $\sim 0.6~\mu\rm m$. This deviation is very sensitive to 
optical alignment and we attribute this to the residual influence of interference between diagonally separated beams in each unit cell. 
The lifetime of atoms in the traps is measured with all light switched off except for the 780~nm trapping light. Measured $1/e$ lifetimes range from 3.7 to 11~s for the different sites. The lifetime limit due to collisions with untrapped background gas was measured to be $\sim 20~\rm s$ using a several mK deep red detuned single beam trap formed with 1040 nm light. 
We attribute the shorter lifetimes in the GBA to the lower trap depth, and possibly imperfections in the trap potentials. 
As with the deviations in the atomic centroids we find the lifetimes to be very sensitive to adjustment 
of the  alignment in the array forming optics. Due to the sensitivity to optical alignment we have studied the performance of the full-incoherent array as described in the following.

\subsection{Full-incoherent array}

\begin{figure}[!t]
 \centering
   \includegraphics[width=8.7cm]{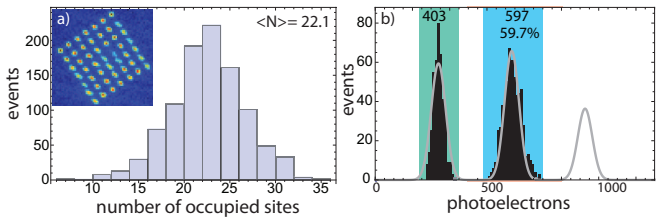}
\caption{(color online)  Atom trapping in a 49 site full-incoherent array.  Distribution of number of occupied sites with fluorescence image (a). The fluorescence image is an average over 100 atom loading shots, with noise filtered by performing a principal component analysis on the data set and combining the 49 components with the highest eigenvalues to form the image.
The histogram from one of the sites (b) shows a 59.7\% single atom loading rate. The solid line is a Poissonian model fitted to the 1 atom peak.   }
\label{fig.49site}
\end{figure}

Atom trapping in the 49 site full-incoherent array is shown in Fig. \ref{fig.49site}. For these experiments we used two separate 780 nm laser systems: the frequency doubled 1560 nm source used for the half-incoherent array experiments and a 
single frequency Ti:Sa laser also operating near 780 nm. The two laser systems were adjusted to have wavelengths within 1 nm of each other. 
The optical efficiency for the setup in Fig. \ref{fig.GBAsetupFI} from the fiber ends to the atoms including the array generation and subsequent relay and focusing optics is about 60\%. The trap depth is given by  Eq. (\ref{eq.gbarrayt}).  With a power of 2.5 W out of each fiber we projected a total of 3 W onto the atoms giving $47~\rm mW$ in each of the 64 beams. This resulted in a trapping potential of   
 $\sim 570~\mu$K in an array with $d=3.8~\mu\rm m$ and $s=2.1$.  Atom trapping was observed with the potential as low as 
 $\sim 340~\mu$K. The data presented below were all taken with  $\sim 570~\mu$K deep traps.  Figure \ref{fig.49site} shows a fluorescence image of the array with a single atom loading histogram. The average atom lifetime in the array was about 1.5 s.
Single atom exposure parameters were the same as for the half-incoherent array.

We see that the average number of occupied sites in each loading event is approximately $22/49=45\%$ of the array size.  Single atoms are loaded into all sites, although about 5 sites have substantially lower loading rates than the average and about 5 sites have loading rates above 60 \%. 
We believe the loading rate could be further improved in all sites using the technique of repulsive light assisted collisions \cite{Grunzweig2010}. The regularity of the atomic positions is significantly better in this array than in the half-incoherent method. Fitting the fluorescence image to a regular grid we find the average deviation of the atomic centroids from a regular grid is about $0.35 ~\mu\rm m$ which is close to half of the deviation seen in the half-incoherent array.

\begin{figure}[!t]
 \centering
   \includegraphics[width=7.5cm]{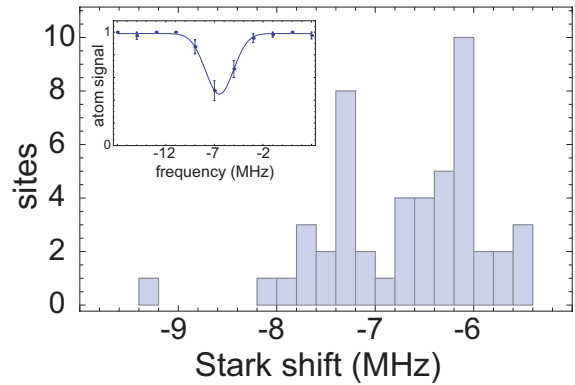}
\caption{(color online) Stark shift of the  $6s_{1/2},f=4\leftrightarrow 6p_{3/2}, f=5$ transition  in the 49 site array. The inset shows an atom blow away curve as the laser frequency is scanned.  }
\label{fig.blowaway}
\end{figure}

An important consideration for qubit experiments is that trap induced Stark shifts on atomic transitions used for state control are uniform across the array. The trapping sites in the GBA are not perfectly dark; there is a non-zero intensity $I_c$  at the center of each trap. This is intentional and is beneficial for correct magic trapping of ground-Rydberg state transitions\cite{SZhang2011}. We have verified the uniformity of  the intensity $I_c$  at  each site by
scanning across the  $6s_{1/2},f=4\leftrightarrow 6p_{3/2}, f=5$ transition with a $5~\mu\rm s$ pulse from a  single unbalanced beam to blow away the atoms. The data  in Fig. \ref{fig.blowaway} show a mean transition Stark shift relative to the value for an atom outside the lattice of $-6.7~\rm MHz$ with a standard deviation of $0.8~\rm  MHz.$  The typical fractional deviation of the Stark shift is thus $0.8 /6.7=12\%.$ 
From Eqs. (\ref{eq.GBAIc2},\ref{eq.gbarrayt}) the intensity at trap center is related to the effective trapping intensity by 
$I_c/I_t = 2 e^{-s^2/2} (1-2 e^{-s^2/2})$, giving $I_c/I_t=0.17$ at $s=2.1$. The average light shift of the ground state at trap center is thus $0.17\times 570~\mu{\rm K} = 97~\mu\rm K$ with a standard deviation of about $10~\mu\rm K$. 

The site to site shifts may come into play when we consider dephasing of qubits encoded in the $f=3, 4$ hyperfine clock states which have a transition frequency of 0.0092 THz. At our detuning of 72 nm (32.5 THz) from the Cs D2 transition the standard deviation of 
the clock frequency across the array due to trap induced Stark shifts is approximately $10~\mu{\rm K} \times \frac{.0092}{.0092 +32.5 }=0.0028~\mu\rm K$ or
$\sim 60~\rm Hz$. If left uncompensated this would cause dephasing of qubits in different sites on a time scale of a few ms.
With good laser stabilization these shifts will be static and it is possible to keep track of them. Alternatively 
much longer coherence times should be possible using additional plane wave optical fields to compensate the differential Stark shift\cite{Radnaev2010}.

\section{Conclusions}

\label{sec.conclusion}

In summary we have designed and demonstrated a novel type of two-dimensional  optical trap array using weakly overlapping Gaussian beams. Single Cs atoms are loaded into the array with approximately 45\% average filling factor. High fidelity detection of single atoms is achieved using fluorescence imaging. The moderately large spacing and blue detuned character of the array make it well suited for demonstrating quantum gates with Rydberg state mediated interactions. Results of quantum
gate experiments in the array using microwave and optical fields for state control will be reported elsewhere.

This work was supported by the IARPA MQCO program
through ARO contract W911NF-10-1-0347 and by
DARPA.


\end{document}